\documentclass[manuscript,screen]{acmart}
\AtBeginDocument{%
  }

\setcopyright{acmlicensed}
\copyrightyear{2026}
\acmYear{2026}
\acmDOI{XXXXXXX.XXXXXXX}
\acmISBN{978-1-4503-XXXX-X/2018/06}

\renewcommand\footnotetextcopyrightpermission[1]{} 

\makeatletter
\let\@authorsaddresses\@empty
\makeatother

\begin{document}

\title{Measuring Perceptions of Fairness in AI Systems: The Effects of Infra-marginality}

\author{Schrasing Tong}
\email{st9@mit.edu}
\affiliation{%
 \institution{Massachusetts Institute of Technology}
 \city{Cambridge}
 \state{Massachusetts}
 \country{United States}}

\author{Minseok Jung}
\affiliation{%
 \institution{Massachusetts Institute of Technology}
 \city{Cambridge}
 \state{Massachusetts}
 \country{United States}}

\author{Ilaria Liccardi}
\affiliation{%
 \institution{Massachusetts Institute of Technology}
 \city{Cambridge}
 \state{Massachusetts}
 \country{United States}}

\author{Lalana Kagal}
\affiliation{%
 \institution{Massachusetts Institute of Technology}
 \city{Cambridge}
 \state{Massachusetts}
 \country{United States}}

\renewcommand{\shortauthors}{Tong et al.}

\begin{abstract}
Differences in data distributions between demographic groups, known as the problem of infra-marginality, complicate how people evaluate fairness in machine learning models.
We present a user study with 85 participants in a hypothetical medical decision-making scenario to examine two treatments: group-specific model performance and training data availability.
Our results show that participants did not equate fairness with simple statistical parity.
When group-specific performances were equal or unavailable, participants preferred models that produced equal outcomes; when performances differed, especially in ways consistent with data imbalances, they judged models that preserved those differences as more fair.
These findings highlight that fairness judgments are shaped not only by outcomes, but also by beliefs about the causes of disparities.
We discuss implications for popular group fairness definitions and system design, arguing that accounting for distributional context is critical to aligning algorithmic fairness metrics with human expectations in real-world applications.
\end{abstract}



\maketitle

\section{Introduction}
The predictions or decisions of a machine learning model may be systematically inaccurate or unfair due to the data it was trained on or the way the algorithm was designed.
These cases often tend to replicate and perpetuate existing inequalities, which can result in significant consequences in applications where the model makes important decisions, such as in hiring~\cite{carey2016companies}, lending~\cite{abellan2017comparative}, and criminal justice~\cite{angwin2016machine}.
To address this problem, researchers have proposed many different definitions of fairness~\cite{berk2021fairness,chouldechova2017fair,corbett2017algorithmic,dwork2012fairness}.
The most widely-used approach in practice relies on sensitive attributes such as race, gender, or age, and resembles constraints between different groups in the population - known as \textbf{group fairness}.
However, there is a lack of consensus on which group fairness definition, for example demographic parity or equal opportunity~\cite{hardt2016equality}, is the most fair or applicable in a given scenario~\cite{gajane2017formalizing} and recent studies have shown that certain definitions could not co-exist~\cite{kleinberg2016inherent}.

Since fairness definitions ultimately seek to capture societal beliefs of fairness in the given scenario, we believe that an empirical approach through user studies would provide precious insights to the ongoing debate. 
Prior works in this area have mostly focused on which fairness definitions are more preferred~\cite{saxena2019fairness,harrison2020empirical} or how other factors influence human perceptions of fairness~\cite{lee2018understanding,woodruff2018qualitative,yurrita2023disentangling}.
In this study, we revisit the most basic, but often overlooked, concept of group fairness: the equality of a certain statistical measure, for example accuracy or false positive rate for recidivism between the groups. 
In practice, sensitive attribute groups often differ in their base rates: the underlying prevalence of the true outcome, for example the proportion of individuals in each group who actually re-offend.
Because fairness metrics such as accuracy or false positive rate depend mathematically on these base rates, they will typically differ across groups even when a single, jointly trained model is used~\cite{corbett2018measure}.
This phenomenon, known as the \textbf{infra-marginality} problem, highlights that disparities in fairness metrics may stem from legitimate base rate differences rather than from harmful discrimination. 
Since the true data distribution is rarely known and base rates are estimated from empirical samples, distinguishing unfair treatment from distributional variation requires careful data analysis and domain knowledge.
If handled incorrectly, fairness interventions can backfire, ultimately harming the very populations they aim to protect. 
For example, forcing defendants of all races in the infamous COMPAS dataset~\cite{angwin2016machine} to have the same predicted recidivism rate would ignore base rates and lead to biased parole decisions. 

To investigate how awareness of infra-marginality shapes perceptions of fairness, we conducted a user study with 85 participants using Qualtrics.
Participants were presented with hypothetical scenarios in which a model was trained on medical data from two racial groups. 
We designed two treatments that varied the information typically revealed by exploratory data analysis (EDA): (1) the performance of models trained separately for each racial group (group-specific performance) and (2) the relative amount of training data available for each race. 
While group-specific performance is not itself a base rate, it reflects how distributional differences between groups, such as different base rates or varying task difficulty, manifest in model outcomes. 
Participants were then asked to rate the fairness of three candidate models using a 7-point Likert scale: (1) a model that equalized performance to match the higher-performing group, testing whether fairness is seen as maximizing outcomes while ensuring parity (2) a model that equalized performance at the average of the two groups, testing whether fairness is perceived as an equal compromise while ensuring parity, and (3) a model that preserved each group’s observed performance, testing whether fairness should respect existing differences when they are viewed as justified.

Our findings show that perceptions of fairness hinge on how participants interpret performance differences between groups.
When group performances were equal or not provided, participants favored models that produced equal accuracy across groups, aligning with standard group fairness definitions. 
In contrast, when performance differences were visible, participants often regarded them as legitimate and expected the joint model to preserve those differences, consistent with an infra-marginality perspective. 
Participants also recognized the role of training data: disparities were considered more acceptable when attributed to inherent task difficulty (reflecting underlying base rate differences), but less acceptable when attributed to biased or imbalanced data collection.

Since base rate and distributional differences are unavoidable in real-world datasets, our findings highlight the need to reconsider how fairness is formally defined in practice.
Widely adopted group fairness metrics, such as equalized error rates, implicitly assume that parity across groups is always the desirable outcome.
However, our study shows that participants often judged parity as unfair when disparities reflected legitimate distributional differences, and conversely viewed preserving such disparities as fairer.
This reveals a critical gap between formal fairness definitions and human perceptions: while metrics emphasize statistical equality, people’s judgments depend on both the magnitude and the perceived causes of disparities.
By examining fairness perceptions through the lens of infra-marginality, we emphasize the importance of developing fairness frameworks that not only satisfy technical criteria but also align with human reasoning, social legitimacy, and the realities of heterogeneous data.

\section{Related Research}
As machine learning became an integral part of various aspects of our lives, fairness has emerged as a critical concern due to the tendency for models to inherit biases in the training data. 
In this context, fairness can be defined as equitable treatment of individuals across different demographic groups, ensuring that algorithmic decisions do not lead to biased or discriminatory outcomes.
Most mathematical definitions revolve around constraints on conditional probabilistic statements between the groups; more advanced and granular definitions such as individual fairness~\cite{dwork2012fairness} remain less popular in practice as they are harder to achieve and justify.
Despite best effort attempts to capture societal expectations, fairness definitions also receive critique in certain scenarios, for example the problem of infra-marginality that has been known to complicate tests for discrimination~\cite{simoiu2017problem}.
To illustrate why we believe user perceptions play a critical role in addressing the issue, we summarize and expand upon the analysis presented by~\cite{corbett2018measure} on the COMPAS dataset. 

The COMPAS dataset contains two groups of defendants, Black and White, with different underlying data distributions and numbers of samples. 
In particular, 21\% of Black defendants and 12\% of White defendants were rearrested, reflecting differences in the base rates of recidivism. 
Some may argue that these disparities are unjust and reflect dataset bias, for example, due to limited or unrepresentative samples, unobserved crimes, or differential policing. 
However, these base rate differences directly affect the calculation of group fairness metrics such as accuracy and false positive rate. 
If the observed differences are legitimate, then forcing identical thresholds across groups, ignoring the underlying base rates, can have harmful consequences, either by imposing harsher parole decisions on some defendants or by releasing higher risk individuals. 
Complicating matters, there are many statistical measures that could be used to define fairness~\cite{verma2018fairness}, and achieving equality on one often worsens disparities on another. 
This example illustrates how infra-marginality arises in practice: disparities in fairness metrics may reflect legitimate base rate differences rather than bias. 
Understanding user perceptions of fairness can help guide how such distributional differences should inform model design and fairness decisions.

There exists significant research on how people perceive an AI model~\cite{kapania2022because,cheng2023overcoming,kim2023one} as well as how transparency and explainability~\cite{panigutti2022understanding,ma2023should} help users understand and trust model predictions and decisions. 
As fairness issues in machine learning become increasingly prevalent, many researchers have taken a human-centric approach through user studies. 
While some focused on specific applications such as targeted advertisement~\cite{plane2017exploring}, recommendation systems~\cite{sonboli2021fairness}, and AI as a service~\cite{lewicki2023out}, the majority of prior works revolve around the traditional fairness setting, where a model makes decisions on subjects from multiple groups. 
Several studies~\cite{saxena2019fairness,srivastava2019mathematical,harrison2020empirical,alkhathlan2024balancing} have attempted to identify the most appropriate fairness definition in given scenarios to overcome the lack of consensus on which definitions are the most fair. 
By quantifying and aggregating user ratings to proposed definitions, these works provide valuable comparisons that guide future model training. 
Our work is heavily inspired by these studies and tackles the problem of different data distributions between groups, which also has no purely technical solution and traces back to user assumptions on the root cause of the problem. 
Other research has also combined model explanations and user fairness perceptions~\cite{dodge2019explaining,schoeffer2022there,binns2018s,van2021effect,yurrita2023disentangling} to see if people could detect bias or discrimination in trained models. 
Despite scientists' best efforts, fairness remains a complicated issue and studies have shown that user perceptions depend on many factors, assumptions, and other details~\cite{grgic2018human,kasinidou2021agree,haider2024crowdsourced,meyer2025perceptions}.
As a result, recent works have proposed frameworks to integrate user perceptions into the training process~\cite{ahnert2021fairceptron,cheng2021soliciting,richardson2021towards}.

Another approach on studying user perceptions of fairness centers more on how humans perceive algorithms and the interaction between users and the trained models~\cite{echterhoff2022ai,boonprakong2023bias,yuan2023contextualizing,ghasemaghaei2024understanding,lima2025laypeople}.
Instead of trying to come up with the most applicable definition of fairness, these studies generate insights on perceptions of automated decision making~\cite{araujo2020ai,tolmeijer2022capable}, ethics~\cite{lima2021human}, and how the process and results are conveyed~\cite{lee2017algorithmic,lee2019procedural}.
Studies have shown that the demographics~\cite{jakesch2022different,pierson2017gender,grgic2020dimensions} and the past experiences~\cite{gemalmaz2022understanding} of participants would influence their perceptions of fairness, especially if they feel like they were subject to bias in the past. 
Furthermore, research has shown that sometimes user perceptions also include inherent bias, such as rating algorithms as more fair when decisions favor themselves~\cite{wang2020factors} or appealing to authority for human decisions yet seeking objectivity for algorithmic ones~\cite{lee2018understanding}. 
In general, understanding these human factors deeply may require interviews~\cite{woodruff2018qualitative} rather than surveys and they provide beneficial insights on the variety of factors and potential limitations of human-centric approaches to fairness. 

\section{User Study}
We designed a user study to investigate how users perceive fairness under infra-marginality, where disparities in model outcomes may arise from legitimate differences in underlying group distributions. 
Our study simulates the type of information available to a machine learning practitioner before deciding whether to deploy a trained algorithm. 
Unlike prior work on user perceptions of fairness~\cite{saxena2019fairness,harrison2020empirical}, which primarily evaluates which fairness definitions users prefer, our study emphasizes the role of distributional differences in shaping fairness judgments. 
Specifically, we aim to address two research questions:
\begin{itemize}
    \item How do users evaluate the fairness of a model given data on underlying distributional differences between groups? 
    \item How does the relative availability of training data influence users’ assumptions about these differences and, in turn, their fairness judgments?
\end{itemize}

\subsection{Study Design}
We conducted an online survey with 28 questions to investigate two key factors shaping user perceptions of fairness under the infra-marginality problem: distributional differences between groups (approximated here by group-specific performances, varied across 7 instances) and relative training data availability (varied across 4 instances).
Group-specific performances provide participants with prior knowledge about differences in model outcomes across groups, directly shaping fairness perceptions since many fairness metrics are defined in terms of equalized performance. 
Data availability serves as a proxy for whether such differences are interpreted as legitimate, reflecting base rates or task difficulty, or as biased, resulting from skewed data collection. 
This factor is only meaningful in interaction with group-specific performances, since data quantity alone does not create fairness concerns when distributions are identical.
The survey scenario was framed around an AI-driven cancer prediction system trained on two groups, labeled Race A and Race B. 
Each question presented a unique combination of the two treatment factors. 
To keep the study tractable and accessible to participants without technical expertise, we simplified aspects of exploratory data analysis (EDA) that are less central to fairness reasoning, such as model architecture, feature correlations, training details, and error decomposition.
For each question, participants were shown the prediction performances of three hypothetical algorithms trained on both groups. 
They were asked to rate the fairness of the algorithms on a 7-point Likert scale (1 = most unfair, 7 = most fair). 
This design enables us to compare participants’ relative preferences for different fairness outcomes and to observe how fairness perceptions shift depending on the treatments.
Example 3.1 shows an example question. 

\begin{example}
A hospital is using an AI driven cancer prediction system to detect cancer in patients of different races.

A model trained only on Race A has 88\% accuracy whereas the model trained only on Race B has 78\% accuracy. The data distribution of the population is unknown but an exploratory data analysis shows that there exists 3x more training data for Race A than for Race B.

Given this information to what extent would you rate the following models as fair/unfair?

Option A: The final system accurately identifies cancer in both Race A and Race B 88\% of the time.

Option B: The final system accurately identifies cancer in both Race A and Race B 83\% of the time.

Option C: The final system accurately identifies cancer in Race A 88\% of the time whereas it accurately identifies cancer in Race B 78\% of the time.
\end{example}

\subsection{Procedure}
We directed participants to an online survey hosted by Qualtrics via a link. 
Since there were only 28 questions, all the participants belonged to one group and responded to the same set of questions.
To prevent survey bias, the order that the questions appeared was randomized for each participant and revisiting previous questions was not allowed. 
The prediction accuracy of the models throughout the survey had been perturbed by a small margin, with changes emphasized in bold, to prevent participants from mistaking them for previous questions; the relative differences between the two groups remained the same. 
 
\subsection{Operationalization}
We sought to accurately replicate the problem of infra-marginality in a practical machine learning setting while simplifying some concepts to make the study non-expert friendly. 

\paragraph{The Scenario}
We picked a medical AI example due to the serious consequences that can arise from a biased model. 
The setting resembles one commonly seen in fairness literature, where researchers analyze whether a model that is trained on two groups performs fairly. 
We chose one of the most studied sensitive attributes, race, and used the more abstract Race A and Race B instead of Caucasians, Asians, or Black to prevent associations that may introduce societal bias. 
To ensure that participants who do not have a medical background can understand the questions and implications, we purposefully did not use any medical jargon or domain-specific examples; one can replace cancer with credit worthiness without compromising the setting. 

\paragraph{Treatment on group-specific performance}
Differences in data distributions are the fundamental driver of the infra-marginality problem. 
If the two groups were very similar, the resulting model would likely appear fair across both. 
However, conveying such distributional differences to participants without deep ML knowledge is challenging. 
In our cancer prediction setting, the natural base rate would be the cancer prevalence in each group. 
While base rates are often used to illustrate infra-marginality, presenting them directly risks confusion or misinterpretation by participants. 
For example, in our cancer prediction scenario, participants would need to recognize that if the cancer prevalence in one group is much lower, then even a trivial model that always predicts "no cancer" can achieve high accuracy. 
In other words, interpreting base rates requires an understanding of how class imbalance affects achievable accuracy. 
Instead, we operationalized infra-marginality using group-specific model accuracies. 
Accuracy incorporates the effects of the underlying distributional differences while aligning more closely with widely used fairness definitions that compare predictive performance across groups.
This choice allows us to communicate the idea of uneven task difficulty in a way that is both accessible to participants and meaningful for fairness evaluation. 
Ultimately, both base rates and accuracies highlight the same underlying issue, but accuracy offers a clearer and less assumption-dependent representation for our user study.
In the study, we introduced 7 different instances for group-specific performances: 1) no information, 2) Race A and Race B have the same accuracy of 90\%, 3) Race A and Race B have the same accuracy of 70\%, 4) Race A has a 10\% higher accuracy than Race B, 95\% vs 85\%, 5) Race A has a 10\% higher accuracy than Race B, 75\% vs 65\%, 6) Race A has a 10\% lower accuracy than Race B, 85\% vs 95\%, and 7) Race A has a 10\% lower accuracy than Race B, 65\% vs 75\%. 
The 10\% difference between groups indicate a clear disparity in the performances of group-specific models while remaining a possible occurrence in real-world datasets (the COMPAS example in Section 2.2).
The instances contain two sets of numbers that follow the same relative difference as we want to ascertain that fairness perceptions are based on the relative difference in performance and not performance itself. 

\paragraph{Treatment on data availability}
Differences in data availability play a critical role in making fairness judgements for two reasons: 1. The amount of data usually corresponds to the privileged and minority groups in the population and 2. More data generally implies that the model can learn to make better predictions. 
For the infra-marginality problem, it provides important information on whether the performance disparity is legitimate. 
We represented data imbalances with 4 different treatments: The data for Race A and Race B could be 1) unspecified, 2) the same amount, 3) Race A has 3x more data than Race B, and 4) Race A has 20x more data than Race B; Race B having more data than Race A is omitted since the combinations are covered in conjunction with group-specific performances. 
A 3x data difference is usually enough to cause significant performance difference in training but falls under the reasonable population differences between the two groups in many regions. 
On the other hand, a 20x data difference will in most cases warrant special learning algorithms due to the large imbalance and reflects an extreme case that rarely occurs when the data collection process uniformly samples the population. 

\paragraph{Experiment controls}
For both treatments, we included instances where both groups have the same data or performance as baseline comparisons. 
We decided to also include instances where those information are missing since participants might not assume that they were equal when unspecified. 

\paragraph{Models presented in the Options}
For each question, participants were asked to rate their fairness perceptions on three models on a Likert scale from 1 to 7. The three Options are:
\begin{list}{}
    \item Option 1: The model performs equally well for both Races, with the same accuracy as the higher group-specific model.
    \item Option 2: The model performs equally well for both Races, with the same accuracy as the average of the two group-specific models. 
    \item Option 3: The model performs differently for the two Races, the two Races each perform the same as their respective group-specific models. 
\end{list}
When group-specific performances are omitted or equal, the three model options preserve the relative inter-group accuracy differences, ensuring a fair comparison across treatments. 
Each option can be interpreted as fair under different assumptions or beliefs about the source of performance disparities:
Option 1 sets both groups’ performance to match the higher-performing group, achieving accuracy parity~\cite{verma2018fairness} while maximizing overall accuracy.
This aligns with the fairness-performance tradeoffs commonly discussed in fairness literature.
Option 2 also enforces accuracy parity but does so at the average of the two group-specific performances, providing a compromise between groups.
Option 3 preserves the original group-specific performances, reflecting a perspective consistent with infra-marginality: performance disparities are fair if they arise from legitimate distributional differences or task difficulty. 
Examining user perceptions of the 3 Options allows us to answer two important questions to the fairness community: 1. Under infra-marginality, do users view striving for strict performance parity as fair? 2. How does this change when assumptions on underlying distributional differences change? 

\subsection{Participants}
The study was approved by our Institutional Review Board and conducted between November 30, 2023, and January 10, 2024. 
We recruited participants via mailing lists in the EECS department at our institution, graduate student dormitories, and LinkedIn groups focused on data science interview practice and knowledge exchange.
We targeted participants with basic familiarity with machine learning, sufficient to understand key concepts such as how training data quantity can affect model performance. 
To capture participants’ self-assessed expertise, we asked them to rate their machine learning knowledge on a 7-point Likert scale (1 = novice, 7 = expert).
Because of this recruitment strategy, our findings are most applicable to individuals with some technical background in computer science or data science, and may not generalize to the broader public. 
Participants who completed the survey were entered into a raffle for Amazon gift cards.

\subsection{Data Validity and Pre-processing}
To ensure participants correctly understood the scenario and questions, we first conducted a pilot study with 8 undergraduate and graduate students in the computer science department. 
After completing the survey, we interviewed them to discuss their interpretations and reasoning. 
This allowed us to verify that participants understood key concepts, such as group-specific performance referring to the accuracy of a model trained on a single group, Race A or B, and evaluated on that same group. 
One valuable insight from the pilot was to add the statement "The data distribution of the population is unknown" when presenting training data composition, which removed potential ambiguity about whether the training data reflected the population.
This effectively prevents participants from considering cases when optimizing for the privileged group can increase overall accuracy. 

To ensure response quality, the final survey included two repeated scenarios. 
Participants whose answers differed substantially on these repeated questions were excluded, as were participants who completed the survey in under 9 minutes or who gave nearly identical scores across most items (standard deviation < 0.3), since these patterns indicate inattention or non-discriminating responses.
In total, 380 participants began the survey, but only 85 met all inclusion criteria, completing all questions, passing the repeated scenario checks, and spending sufficient time engaging with the material; 186 participants did not complete the survey. 
For the repeated scenario check, we removed participants whose responses differed by more than 1.5 points (on a 7-point scale) on average across the six repeated items (two questions × three model options).
All statistical analyses were conducted using Independent T-Tests, as the fairness ratings approximately followed a normal distribution. 
This combination of pilot testing, attention checks, and exclusion criteria ensures that our analyses reflect participants who meaningfully engaged with the survey and understood the presented information.

\section{Results}
\subsection{Participant Demographics}
We first describe participant demographics. 
Of the 85 participants, 42 identified as female, 42 as male, and 1 as non-binary. 
Regarding race/ethnicity, 12 participants identified as Black, 15 as Asian, 4 as Hispanic, and 54 as White. 
Participants’ ages ranged from 21 to 42, with a mean of 29.9 years.
In terms of education, 5 participants had a high school diploma or below, 47 held or were pursuing a college or undergraduate degree, and 33 held or were pursuing an advanced or professional degree.
For self-rated machine learning expertise (on a 7-point Likert scale), 6 participants reported low familiarity (ratings 1–3), 27 were moderate (rating 4), and 52 reported higher familiarity (ratings 5–7). 
We did not separate analyses by self-rated expertise due to the potential influence of the Dunning-Kruger effect, where self-assessment may not reliably reflect actual skill.
Overall, these demographics indicate that our findings are most applicable to younger, well-educated individuals with a technical or computer science background, rather than the general population.

\subsection{How Group-Specific Performance Affects Fairness Perceptions}
Figure 1 shows how perceived fairness varies across the seven instances of the group-specific performance treatment, with responses aggregated across data availability. 
Subplots 2–3, 4–5, and 6–7 each represent treatments with the same relative accuracy difference between the two races. 
These side-by-side comparisons indicate that the absolute accuracy values have minimal impact on fairness ratings, supporting the decision to combine instances with the same relative difference in subsequent analyses.
Subplots 2 and 3, where group-specific performances are equal, serve as a control. 
In these conditions, Options 1 and 2 consistently received higher fairness ratings than Option 3. 
Subplots 4–7 correspond to treatments with a 10\% accuracy difference between the groups. 
In these cases, Option 3 (matching the group-specific performances) received the highest mean ratings, which were statistically significant compared to Option 2 and generally higher than Option 1 (p < 0.1 for three of the four subplots).
Subplot 1 shows a control condition with no information about group-specific performances. 
Here, Option 1 received statistically higher fairness ratings than both Options 2 and 3.

\begin{figure}[t]
    \includegraphics[width=\textwidth]{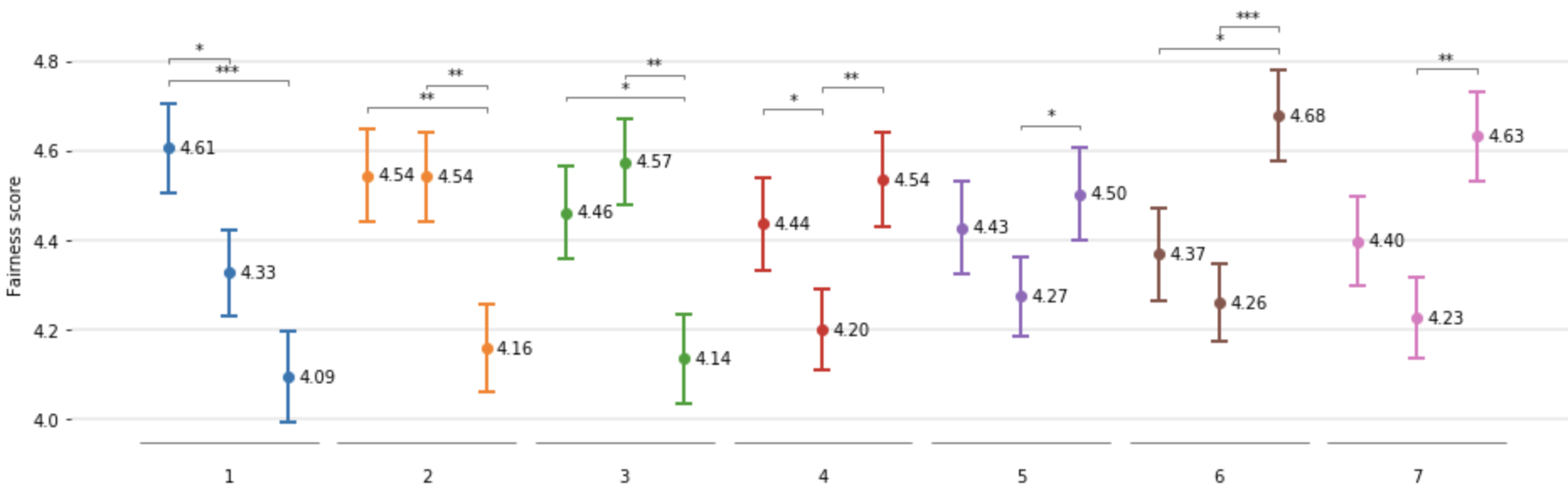}
    \caption{Mean and standard errors of fairness perceptions on the 3 Options for the group-specific performance treatment. Group-specific accuracy denoted as (Race A and Race B) for the 7 subplots are NA/NA, 90/90, 70/70, 95/85, 75/65, 85/95, and 65/75. * signifies p $<$ 0.05, ** signifies p $<$ 0.01, and *** signifies p $<$ 0.001.}
    \label{4-1}
\end{figure}

\subsection{The Effects of Data Balance}
We analyzed how varying the amount of training data for the two races affected perceived fairness. 
Figures 2 and 3 show fairness ratings across the four instances of the data availability treatment, separated by whether Race A had higher or lower group-specific performance than Race B.
For instances with no information on data quantity (subplot 1) and equal amounts of data (subplot 4), results were similar between Figures 2 and 3, reflecting the symmetry between Race A and Race B. 
In these cases, Option 3 received the highest mean fairness ratings, with statistically significant differences observed between Options 2 and 3 when data amounts were equal.
For instances where Race A had 3x or 20x more data than Race B (subplots 2 and 3), the patterns differed between Figures 2 and 3.
Option 3 remained the highest-rated option across all four subplots. 
In Figure 2, where Race A had higher performance, differences between options were not statistically significant. 
In Figure 3, where Race A had lower performance, Option 3 was rated significantly higher than the other options in three out of four comparisons.

\begin{figure}[t]
    \centering
    \includegraphics[width=\textwidth]{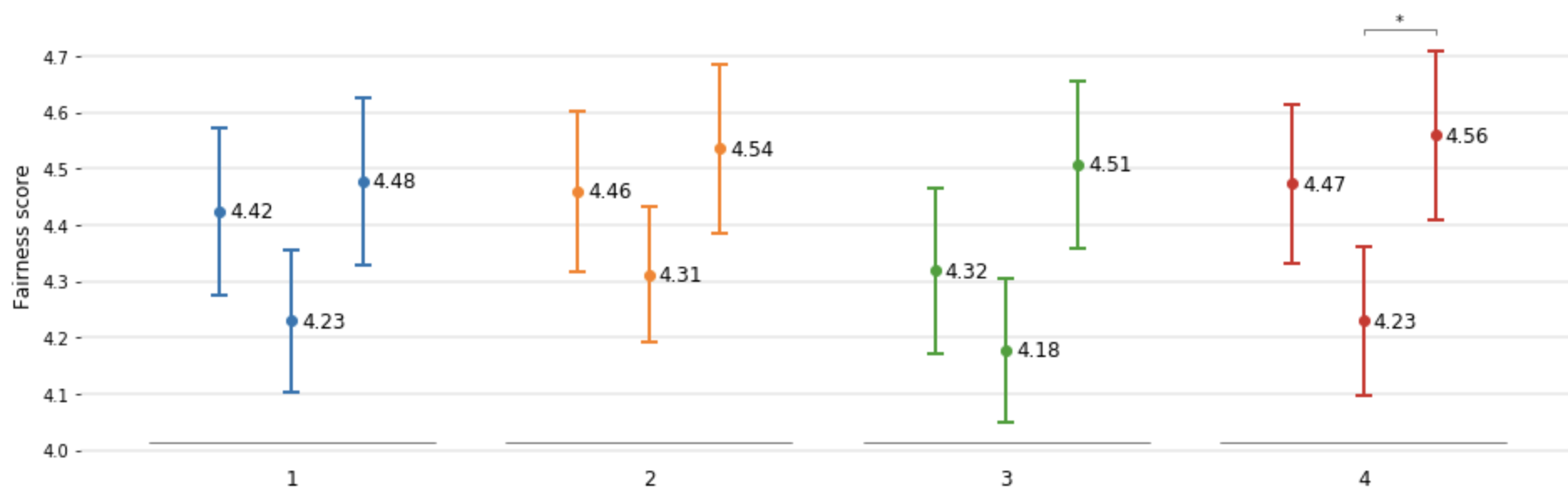}
    \caption{Mean and standard errors of fairness perceptions on the 3 Options when Race A $>$ Race B in group-specific performance. Subplots show data of Race A relative to Race B: no info, 3x, 20x, and 1x respectively. * signifies p $<$ 0.05, ** signifies p $<$ 0.01, and *** signifies p $<$ 0.001.}
    \label{4-2}
\end{figure}

\begin{figure}[t]
    \centering
    \includegraphics[width=\textwidth]{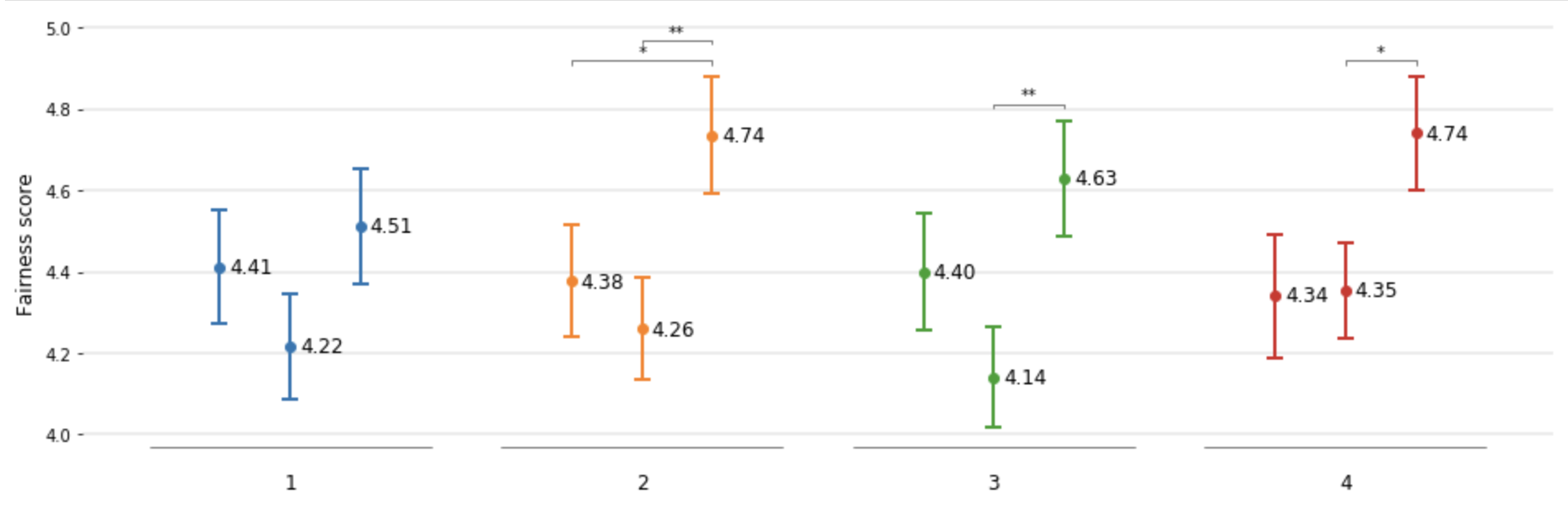}
    \caption{Mean and standard errors of fairness perceptions on the 3 Options when Race A $<$ Race B in group-specific performance. Subplots show data of Race A relative to Race B: no info, 3x, 20x, and 1x respectively. * signifies p $<$ 0.05, ** signifies p $<$ 0.01, and *** signifies p $<$ 0.001.}
    \label{4-3}
\end{figure}

\section{Discussion and Implications}
Our results demonstrate that information about distributional differences between groups, operationalized as group-specific performances, has a strong influence on user perceptions of fairness. 
When group-specific performances were equal or not provided, participants consistently preferred models that treated the groups identically. 
Conversely, when performance differences were present, participants tended to favor options that preserved those differences, reflecting a sensitivity to the underlying variability between groups rather than a blanket desire for parity.
In real-world settings, it is common for distributions and in turn base rates or any statistical measure to differ between groups to some extent. 
Deciding whether to enforce strict parity on a particular fairness definition requires careful consideration. 
Both the magnitude of the differences and, critically, the reasons behind them can justify different approaches: differences arising from intrinsic task difficulty or legitimate distributional variation may warrant preserving disparities, whereas differences resulting from biased sampling or other data limitations may justify interventions to equalize performance.

Our results further indicate that participants integrate information about data availability when forming fairness judgments. 
When a group had higher performance but more training data, participants did not automatically interpret this as fair, whereas higher performance paired with less data was more likely interpreted as a justified outcome of task difficulty. 
This suggests that participants implicitly reason about how underlying data distributions shape model behavior.
In practice, machine learning researchers often account for differences in group data through sampling strategies or re-weighting to reduce bias. 
Our findings suggest that these data-related considerations may also influence end users’ perceptions of fairness, highlighting that assumptions about data quantity and quality are relevant not only for model training but also for communicating fairness to stakeholders. 

Our study suggests that user perceptions of fairness often incorporate a relativity factor, rather than being based solely on absolute differences between groups. 
For example, when participants evaluated Options 1 and 2, consensus was limited, and much of the reasoning referred to how the outcomes compared to the "original" group-specific performances. 
Similarly, in scenarios where group-specific performances differed, choosing Option 1 would have improved performance for one or both groups, but participants often rated Option 3, the option that preserved the original disparities, as fairer.
These patterns indicate that users do not evaluate fairness solely through a utilitarian lens, such as maximizing overall performance, or through strict equality. 
This pattern aligns with well-established concepts in human judgment, such as anchoring and reference dependence, where people evaluate outcomes relative to an initial baseline rather than in absolute terms. 
In the context of fairness, participants appear to treat the original group-specific performances as a reference point, which shapes how they perceive changes introduced by the jointly-trained model.

Overall, these findings highlight a critical gap between formal fairness definitions and how fairness is perceived in practice. 
Widely adopted group fairness metrics, such as equalized error rates, implicitly assume that parity across groups is always the desired outcome. 
Yet, our participants judged parity as unfair when disparities reflected legitimate distributional differences, and conversely viewed preserving those disparities as fairer. 
This matters because in many real-world settings, such as criminal justice, healthcare, and education, algorithmic decisions must not only satisfy technical fairness metrics but also gain legitimacy and trust from affected communities. 
If interventions that enforce parity are perceived as unfair or misaligned with human reasoning, they risk undermining both adoption and public confidence in algorithmic systems. 
Our results therefore suggest that fairness research should move beyond parity-focused interventions to frameworks that incorporate both statistical rigor and sensitivity to the causes and contexts of group disparities.

This paper has primarily focused on group fairness and the infra-marginality problem, but several of our findings extend to stronger notions of fairness, such as subgroup fairness\cite{kearns2018preventing} and individual fairness\cite{dwork2012fairness}. 
A well-known limitation of group fairness is that evaluating only one sensitive attribute often underestimates the extent of unfairness. 
For example, even if Race A and Race B achieve equal performance, this does not guarantee that the four intersectional groups formed by Race and Gender will also perform equally well. 
Extending this logic, one can continue subdividing the population into finer subgroups, eventually reaching the individual level.
Our results suggest that information about differences in data distributions meaningfully shapes fairness perceptions, and that strict equality across groups may not always be perceived as the most fair. 
As we consider increasingly fine-grained subgroups, such distributional differences are likely to grow, making rigid parity constraints less realistic or even counterproductive. 
This highlights the need for fairness approaches that can flexibly account for distributional context rather than enforcing equality at all costs.
Individual fairness addresses this challenge by requiring that similar individuals receive similar outcomes, operationalized through a predefined similarity metric. 
However, defining and agreeing upon such a metric in practice remains a central open problem.

\section{Conclusion}
This paper examined how users perceive fairness under infra-marginality, where disparities in model outcomes may arise from legitimate differences in underlying group distributions.
Through a user study with 85 participants, we investigated two treatments: group-specific performance and data availability.
Our key findings include: 
\begin{itemize}
    \item \textbf{Perceived fairness depends on group-specific performance baselines.} When group-specific performances were equal or unavailable, participants preferred models that equalized outcomes. When performances differed, they expected the final model to preserve those differences. 
    \item \textbf{Data availability shapes fairness reasoning.} Participants recognized that unequal training data could drive performance disparities, and incorporated this into their fairness judgments on task difficulty versus bias.
    \item \textbf{Fairness is evaluated relative to baselines.} Participants often compared candidate models to the original group-specific performances, suggesting an anchoring effect rather than purely outcome-based reasoning. 
    \item \textbf{Parity-based group fairness metrics diverge from user expectations.} Widely used group fairness definitions, such as equalized error rates, may conflict with how people judge fairness when disparities arise from distributional differences rather than algorithmic bias.
\end{itemize}
These findings show that fairness judgments are not based solely on parity across groups, but also on the distributional context and the perceived legitimacy of disparities. 
Recognizing this matters: fairness interventions that override what users see as justified differences may undermine trust in algorithmic systems, while ignoring disparities rooted in structural inequities risks reinforcing injustice. 
By situating fairness perceptions within the infra-marginality problem, this work highlights the need for fairness definitions and interventions that account for real-world heterogeneity and user reasoning, bridging formal fairness metrics with human expectations.


\bibliographystyle{ACM-Reference-Format}
\bibliography{sample-base}

@String{Computing = "Computing" }

@String{Computer = "{IEEE} Computer" }

@String{Springer = "Springer-Verlag" }

@inproceedings{wang2020factors,
  title={Factors influencing perceived fairness in algorithmic decision-making: Algorithm outcomes, development procedures, and individual differences},
  author={Wang, Ruotong and Harper, F Maxwell and Zhu, Haiyi},
  booktitle={Proceedings of the 2020 CHI Conference on Human Factors in Computing Systems},
  pages={1--14},
  year={2020}
}

@inproceedings{woodruff2018qualitative,
  title={A qualitative exploration of perceptions of algorithmic fairness},
  author={Woodruff, Allison and Fox, Sarah E and Rousso-Schindler, Steven and Warshaw, Jeffrey},
  booktitle={Proceedings of the 2018 chi conference on human factors in computing systems},
  pages={1--14},
  year={2018}
}

@article{lee2018understanding,
  title={Understanding perception of algorithmic decisions: Fairness, trust, and emotion in response to algorithmic management},
  author={Lee, Min Kyung},
  journal={Big Data \& Society},
  volume={5},
  number={1},
  pages={2053951718756684},
  year={2018},
  publisher={SAGE Publications Sage UK: London, England}
}

@article{jakesch2022different,
  title={How Different Groups Prioritize Ethical Values for Responsible AI},
  author={Jakesch, Maurice and Bu{\c{c}}inca, Zana and Amershi, Saleema and Olteanu, Alexandra},
  journal={arXiv preprint arXiv:2205.07722},
  year={2022}
}

@inproceedings{saxena2019fairness,
  title={How do fairness definitions fare? {Examining} public attitudes towards algorithmic definitions of fairness},
  author={Saxena, Nripsuta Ani and Huang, Karen and DeFilippis, Evan and Radanovic, Goran and Parkes, David C and Liu, Yang},
  booktitle={Proceedings of the 2019 AAAI/ACM Conference on AI, Ethics, and Society},
  pages={99--106},
  year={2019}
}

@inproceedings{dodge2019explaining,
  title={Explaining models: an empirical study of how explanations impact fairness judgment},
  author={Dodge, Jonathan and Liao, Q Vera and Zhang, Yunfeng and Bellamy, Rachel KE and Dugan, Casey},
  booktitle={Proceedings of the 24th international conference on intelligent user interfaces},
  pages={275--285},
  year={2019}
}

@inproceedings{binns2018s,
  title={'{It's} Reducing a Human Being to a Percentage' Perceptions of Justice in Algorithmic Decisions},
  author={Binns, Reuben and Van Kleek, Max and Veale, Michael and Lyngs, Ulrik and Zhao, Jun and Shadbolt, Nigel},
  booktitle={Proceedings of the 2018 Chi conference on human factors in computing systems},
  pages={1--14},
  year={2018}
}

@inproceedings{srivastava2019mathematical,
  title={Mathematical notions vs. human perception of fairness: A descriptive approach to fairness for machine learning},
  author={Srivastava, Megha and Heidari, Hoda and Krause, Andreas},
  booktitle={Proceedings of the 25th ACM SIGKDD international conference on knowledge discovery \& data mining},
  pages={2459--2468},
  year={2019}
}

@inproceedings{harrison2020empirical,
  title={An empirical study on the perceived fairness of realistic, imperfect machine learning models},
  author={Harrison, Galen and Hanson, Julia and Jacinto, Christine and Ramirez, Julio and Ur, Blase},
  booktitle={Proceedings of the 2020 conference on fairness, accountability, and transparency},
  pages={392--402},
  year={2020}
}

@inproceedings{van2021effect,
  title={Effect of information presentation on fairness perceptions of machine learning predictors},
  author={Van Berkel, Niels and Goncalves, Jorge and Russo, Daniel and Hosio, Simo and Skov, Mikael B},
  booktitle={Proceedings of the 2021 CHI Conference on Human Factors in Computing Systems},
  pages={1--13},
  year={2021}
}

@inproceedings{kasinidou2021agree,
  title={I agree with the decision, but they didn't deserve this: Future Developers' Perception of Fairness in Algorithmic Decisions},
  author={Kasinidou, Maria and Kleanthous, Styliani and Barlas, P{\i}nar and Otterbacher, Jahna},
  booktitle={Proceedings of the 2021 acm conference on fairness, accountability, and transparency},
  pages={690--700},
  year={2021}
}

@inproceedings{lima2021human,
  title={Human perceptions on moral responsibility of {AI}: A case study in {AI}-assisted bail decision-making},
  author={Lima, Gabriel and Grgi{\'c}-Hla{\v{c}}a, Nina and Cha, Meeyoung},
  booktitle={Proceedings of the 2021 CHI Conference on Human Factors in Computing Systems},
  pages={1--17},
  year={2021}
}

@article{grgic2020dimensions,
  title={Dimensions of diversity in human perceptions of algorithmic fairness},
  author={Grgi{\'c}-Hla{\v{c}}a, Nina and Weller, Adrian and Redmiles, Elissa M},
  journal={arXiv preprint arXiv:2005.00808},
  year={2020}
}

@article{schoeffer2022there,
  title={" {There} Is Not Enough Information": On the Effects of Explanations on Perceptions of Informational Fairness and Trustworthiness in Automated Decision-Making},
  author={Schoeffer, Jakob and Kuehl, Niklas and Machowski, Yvette},
  journal={arXiv preprint arXiv:2205.05758},
  year={2022}
}

@inproceedings{ahnert2021fairceptron,
  title={The FairCeptron: A framework for measuring human perceptions of algorithmic fairness},
  author={Ahnert, Georg and Smirnov, Ivan and Lemmerich, Florian and Wagner, Claudia and Strohmaier, Markus},
  booktitle={Adjunct Proceedings of the 29th ACM Conference on User Modeling, Adaptation and Personalization},
  pages={401--403},
  year={2021}
}

@inproceedings{gemalmaz2022understanding,
  title={Understanding Decision Subjects' Fairness Perceptions and Retention in Repeated Interactions with {AI}-Based Decision Systems},
  author={Gemalmaz, Meric Altug and Yin, Ming},
  booktitle={Proceedings of the 2022 AAAI/ACM Conference on AI, Ethics, and Society},
  pages={295--306},
  year={2022}
}

@inproceedings{cheng2021soliciting,
  title={Soliciting stakeholders’ fairness notions in child maltreatment predictive systems},
  author={Cheng, Hao-Fei and Stapleton, Logan and Wang, Ruiqi and Bullock, Paige and Chouldechova, Alexandra and Wu, Zhiwei Steven Steven and Zhu, Haiyi},
  booktitle={Proceedings of the 2021 CHI Conference on Human Factors in Computing Systems},
  pages={1--17},
  year={2021}
}

@inproceedings{plane2017exploring,
  title={Exploring user perceptions of discrimination in online targeted advertising},
  author={Plane, Angelisa C and Redmiles, Elissa M and Mazurek, Michelle L and Tschantz, Michael Carl},
  booktitle={26th USENIX Security Symposium (USENIX Security 17)},
  pages={935--951},
  year={2017}
}

@inproceedings{sonboli2021fairness,
  title={Fairness and transparency in recommendation: The users’ perspective},
  author={Sonboli, Nasim and Smith, Jessie J and Cabral Berenfus, Florencia and Burke, Robin and Fiesler, Casey},
  booktitle={Proceedings of the 29th ACM Conference on User Modeling, Adaptation and Personalization},
  pages={274--279},
  year={2021}
}

@inproceedings{grgic2018human,
  title={Human perceptions of fairness in algorithmic decision making: A case study of criminal risk prediction},
  author={Grgic-Hlaca, Nina and Redmiles, Elissa M and Gummadi, Krishna P and Weller, Adrian},
  booktitle={Proceedings of the 2018 World Wide Web Conference},
  pages={903--912},
  year={2018}
}

@inproceedings{lee2017algorithmic,
  title={Algorithmic mediation in group decisions: Fairness perceptions of algorithmically mediated vs. discussion-based social division},
  author={Lee, Min Kyung and Baykal, Su},
  booktitle={Proceedings of the 2017 acm conference on computer supported cooperative work and social computing},
  pages={1035--1048},
  year={2017}
}

@article{pierson2017gender,
  title={Gender differences in beliefs about algorithmic fairness},
  author={Pierson, Emma},
  journal={arXiv preprint arXiv:1712.09124},
  year={2017}
}

@article{lee2019procedural,
  title={Procedural justice in algorithmic fairness: Leveraging transparency and outcome control for fair algorithmic mediation},
  author={Lee, Min Kyung and Jain, Anuraag and Cha, Hea Jin and Ojha, Shashank and Kusbit, Daniel},
  journal={Proceedings of the ACM on Human-Computer Interaction},
  volume={3},
  number={CSCW},
  pages={1--26},
  year={2019},
  publisher={ACM New York, NY, USA}
}

@article{araujo2020ai,
  title={In {AI} we trust? Perceptions about automated decision-making by artificial intelligence},
  author={Araujo, Theo and Helberger, Natali and Kruikemeier, Sanne and De Vreese, Claes H},
  journal={AI \& SOCIETY},
  volume={35},
  number={3},
  pages={611--623},
  year={2020},
  publisher={Springer}
}

@article{corbett2018measure,
  title={The measure and mismeasure of fairness: A critical review of fair machine learning},
  author={Corbett-Davies, Sam and Goel, Sharad},
  journal={arXiv preprint arXiv:1808.00023},
  year={2018}
}

@inproceedings{dwork2012fairness,
  title={Fairness through awareness},
  author={Dwork, Cynthia and Hardt, Moritz and Pitassi, Toniann and Reingold, Omer and Zemel, Richard},
  booktitle={Proceedings of the 3rd innovations in theoretical computer science conference},
  pages={214--226},
  year={2012}
}

@misc{angwin2016machine, 
title={Machine bias risk assessments in criminal sentencing.}, 
url={https://www.propublica.org/article/machine-bias-risk-assessments-in-criminal-sentencing}, 
journal={ProPublica}, 
author={Angwin, Julia and Larson, Jeff and Kirchner, Lauren and Mattu, Surya}, 
year={2016}, 
month={May}
}

@article{gajane2017formalizing,
  title={On formalizing fairness in prediction with machine learning},
  author={Gajane, Pratik and Pechenizkiy, Mykola},
  journal={arXiv preprint arXiv:1710.03184},
  year={2017}
}

@article{kleinberg2016inherent,
  title={Inherent trade-offs in the fair determination of risk scores},
  author={Kleinberg, Jon and Mullainathan, Sendhil and Raghavan, Manish},
  journal={arXiv preprint arXiv:1609.05807},
  year={2016}
}

@article{chouldechova2017fair,
  title={Fair prediction with disparate impact: A study of bias in recidivism prediction instruments},
  author={Chouldechova, Alexandra},
  journal={Big data},
  volume={5},
  number={2},
  pages={153--163},
  year={2017},
  publisher={Mary Ann Liebert, Inc. 140 Huguenot Street, 3rd Floor New Rochelle, NY 10801 USA}
}

@inproceedings{corbett2017algorithmic,
  title={Algorithmic decision making and the cost of fairness},
  author={Corbett-Davies, Sam and Pierson, Emma and Feller, Avi and Goel, Sharad and Huq, Aziz},
  booktitle={Proceedings of the 23rd acm sigkdd international conference on knowledge discovery and data mining},
  pages={797--806},
  year={2017}
}

@article{carey2016companies,
  title={How companies are using simulations, competitions, and analytics to hire},
  author={Carey, Dennis and Smith, Matt},
  journal={Harvard Business Review},
  year={2016}
}

@article{abellan2017comparative,
  title={A comparative study on base classifiers in ensemble methods for credit scoring},
  author={Abell{\'a}n, Joaqu{\'\i}n and Castellano, Javier G},
  journal={Expert systems with applications},
  volume={73},
  pages={1--10},
  year={2017},
  publisher={Elsevier}
}

@article{berk2021fairness,
  title={Fairness in criminal justice risk assessments: The state of the art},
  author={Berk, Richard and Heidari, Hoda and Jabbari, Shahin and Kearns, Michael and Roth, Aaron},
  journal={Sociological Methods \& Research},
  volume={50},
  number={1},
  pages={3--44},
  year={2021},
  publisher={Sage Publications Sage CA: Los Angeles, CA}
}

@inproceedings{verma2018fairness,
  title={Fairness definitions explained},
  author={Verma, Sahil and Rubin, Julia},
  booktitle={2018 ieee/acm international workshop on software fairness (fairware)},
  pages={1--7},
  year={2018},
  organization={IEEE}
}

@article{simoiu2017problem,
  title={The problem of infra-marginality in outcome tests for discrimination},
  author={Simoiu, Camelia and Corbett-Davies, Sam and Goel, Sharad},
  journal={The Annals of Applied Statistics},
  volume={11},
  number={3},
  pages={1193--1216},
  year={2017},
  publisher={Institute of Mathematical Statistics}
}

@inproceedings{kearns2018preventing,
  title={Preventing fairness gerrymandering: Auditing and learning for subgroup fairness},
  author={Kearns, Michael and Neel, Seth and Roth, Aaron and Wu, Zhiwei Steven},
  booktitle={International Conference on Machine Learning},
  pages={2564--2572},
  year={2018},
  organization={PMLR}
}

@inproceedings{richardson2021towards,
  title={Towards fairness in practice: A practitioner-oriented rubric for evaluating Fair ML Toolkits},
  author={Richardson, Brianna and Garcia-Gathright, Jean and Way, Samuel F and Thom, Jennifer and Cramer, Henriette},
  booktitle={Proceedings of the 2021 CHI Conference on Human Factors in Computing Systems},
  pages={1--13},
  year={2021}
}

@inproceedings{panigutti2022understanding,
  title={Understanding the impact of explanations on advice-taking: a user study for AI-based clinical Decision Support Systems},
  author={Panigutti, Cecilia and Beretta, Andrea and Giannotti, Fosca and Pedreschi, Dino},
  booktitle={Proceedings of the 2022 CHI Conference on Human Factors in Computing Systems},
  pages={1--9},
  year={2022}
}

@inproceedings{kapania2022because,
  title={” Because AI is 100\% right and safe”: User attitudes and sources of AI authority in India},
  author={Kapania, Shivani and Siy, Oliver and Clapper, Gabe and SP, Azhagu Meena and Sambasivan, Nithya},
  booktitle={Proceedings of the 2022 CHI Conference on Human Factors in Computing Systems},
  pages={1--18},
  year={2022}
}

@inproceedings{tolmeijer2022capable,
  title={Capable but amoral? Comparing AI and human expert collaboration in ethical decision making},
  author={Tolmeijer, Suzanne and Christen, Markus and Kandul, Serhiy and Kneer, Markus and Bernstein, Abraham},
  booktitle={Proceedings of the 2022 CHI Conference on Human Factors in Computing Systems},
  pages={1--17},
  year={2022}
}

@inproceedings{echterhoff2022ai,
  title={AI-moderated decision-making: Capturing and balancing anchoring bias in sequential decision tasks},
  author={Echterhoff, Jessica Maria and Yarmand, Matin and McAuley, Julian},
  booktitle={Proceedings of the 2022 CHI Conference on Human Factors in Computing Systems},
  pages={1--9},
  year={2022}
}

@inproceedings{cheng2023overcoming,
  title={Overcoming Algorithm Aversion: A Comparison between Process and Outcome Control},
  author={Cheng, Lingwei and Chouldechova, Alexandra},
  booktitle={Proceedings of the 2023 CHI Conference on Human Factors in Computing Systems},
  pages={1--27},
  year={2023}
}

@inproceedings{ma2023should,
  title={Who Should I Trust: AI or Myself? Leveraging Human and AI Correctness Likelihood to Promote Appropriate Trust in AI-Assisted Decision-Making},
  author={Ma, Shuai and Lei, Ying and Wang, Xinru and Zheng, Chengbo and Shi, Chuhan and Yin, Ming and Ma, Xiaojuan},
  booktitle={Proceedings of the 2023 CHI Conference on Human Factors in Computing Systems},
  pages={1--19},
  year={2023}
}

@inproceedings{kim2023one,
  title={One AI Does Not Fit All: A Cluster Analysis of the Laypeople’s Perception of AI Roles},
  author={Kim, Taenyun and Molina, Maria D and Rheu, Minjin and Zhan, Emily S and Peng, Wei},
  booktitle={Proceedings of the 2023 CHI Conference on Human Factors in Computing Systems},
  pages={1--20},
  year={2023}
}

@inproceedings{boonprakong2023bias,
  title={Bias-Aware Systems: Exploring Indicators for the Occurrences of Cognitive Biases when Facing Different Opinions},
  author={Boonprakong, Nattapat and Chen, Xiuge and Davey, Catherine and Tag, Benjamin and Dingler, Tilman},
  booktitle={Proceedings of the 2023 CHI Conference on Human Factors in Computing Systems},
  pages={1--19},
  year={2023}
}

@inproceedings{yuan2023contextualizing,
  title={Contextualizing User Perceptions about Biases for Human-Centered Explainable Artificial Intelligence},
  author={Yuan, Chien Wen and Bi, Nanyi and Lin, Ya-Fang and Tseng, Yuen-Hsien},
  booktitle={Proceedings of the 2023 CHI Conference on Human Factors in Computing Systems},
  pages={1--15},
  year={2023}
}

@inproceedings{lewicki2023out,
  title={Out of Context: Investigating the Bias and Fairness Concerns of “Artificial Intelligence as a Service”},
  author={Lewicki, Kornel and Lee, Michelle Seng Ah and Cobbe, Jennifer and Singh, Jatinder},
  booktitle={Proceedings of the 2023 CHI Conference on Human Factors in Computing Systems},
  pages={1--17},
  year={2023}
}

@inproceedings{yurrita2023disentangling,
  title={Disentangling Fairness Perceptions in Algorithmic Decision-Making: the Effects of Explanations, Human Oversight, and Contestability},
  author={Yurrita, Mireia and Draws, Tim and Balayn, Agathe and Murray-Rust, Dave and Tintarev, Nava and Bozzon, Alessandro},
  booktitle={Proceedings of the 2023 CHI Conference on Human Factors in Computing Systems},
  pages={1--21},
  year={2023}
}

@article{hardt2016equality,
  title={Equality of opportunity in supervised learning},
  author={Hardt, Moritz and Price, Eric and Srebro, Nati},
  journal={Advances in neural information processing systems},
  volume={29},
  year={2016}
}

@inproceedings{meyer2025perceptions,
  title={Perceptions of the Fairness Impacts of Multiplicity in Machine Learning},
  author={Meyer, Anna P and Kim, Yea-Seul and D'Antoni, Loris and Albarghouthi, Aws},
  booktitle={Proceedings of the 2025 CHI Conference on Human Factors in Computing Systems},
  pages={1--15},
  year={2025}
}

@article{ghasemaghaei2024understanding,
  title={Understanding how algorithmic injustice leads to making discriminatory decisions: An obedience to authority perspective},
  author={Ghasemaghaei, Maryam and Kordzadeh, Nima},
  journal={Information \& Management},
  volume={61},
  number={2},
  pages={103921},
  year={2024},
  publisher={Elsevier}
}

@inproceedings{haider2024crowdsourced,
  title={Do crowdsourced fairness preferences correlate with risk perceptions?},
  author={Haider, Chowdhury Mohammad Rakin and Clifton, Christopher and Yin, Ming},
  booktitle={Proceedings of the 29th International Conference on Intelligent User Interfaces},
  pages={304--324},
  year={2024}
}

@inproceedings{alkhathlan2024balancing,
  title={Balancing Act: Evaluating People’s Perceptions of Fair Ranking Metrics},
  author={Alkhathlan, Mallak and Cachel, Kathleen and Shrestha, Hilson and Harrison, Lane and Rundensteiner, Elke},
  booktitle={Proceedings of the 2024 ACM Conference on Fairness, Accountability, and Transparency},
  pages={1940--1970},
  year={2024}
}

@article{lima2025laypeople,
  title={Laypeople's Attitudes Towards Fair, Affirmative, and Discriminatory Decision-Making Algorithms},
  author={Lima, Gabriel and Grgi{\'c}-Hla{\v{c}}a, Nina and Langer, Markus and Zou, Yixin},
  journal={arXiv preprint arXiv:2505.07339},
  year={2025}
}


\end{document}